\documentclass[12pt]{article}
\usepackage{amsmath}
\usepackage{amsmath,amsthm}
\usepackage{graphicx}
\begin{document}
\baselineskip=17pt
\pagestyle{empty}
{\author{Alexander Loskutov and Alexei Ryabov\\
{\small Physics Faculty, Moscow State University,
Moscow 119899 Russia}}
\title{Chaotic time-dependent billiards}
\date{ }
\maketitle

\begin{abstract}

A billiard in the form of a stadium with periodically perturbed
boundary is considered. Two types of such billiards
are studied: stadium with strong chaotic properties and a
near-rectangle billiard. Phase portraits of such billiards are
investigated. In the phase plane areas corresponding to
decrease and increase of the velocity of billiard particles are
found. Average velocities of the particle ensemble as functions
of the number of collisions are obtained.

\end{abstract}

Keywords: chaos, billiards, phase portraits

\thispagestyle{empty}

\section{Introduction}

A notion of billiard in physics is known since G.Bikhoff \cite{Birkgof}
who considered a problem concerning the free motion
of a point particle (billiard ball) in some bounded manifold.
So, the billiard dynamical system can be introduced as follows.
A billiard table $Q$ is a Riemannian manifold $M$ with a
piecewise smooth boundary $\partial Q$. The billiard particle freely
moves in $Q$. Reaching the boundary, it is reflected from it
elastically. Thus, the billiard particle moves along geodesic
lines with a constant velocity. In the present article we consider
billiards in Euclidean plane. In this case the angle of incidence of
the particle is always equal to the angle of reflection.

In accordance with the boundary geometry, dynamics of the billiard
particle can be integrable \cite{bMarkaryan2}, completely chaotic
\cite{Sinai3, BnchSinLorenzGas} and, depending on the
initial conditions, regular or chaotic
\cite{bMarkaryan3, BnchEgodicProp1, BnchEgodicProp2, ZaslKriter}.

A natural physical generalization of the billiard problem is perturbation
of the boundary in one or another manner. For the first time, this
problem concerning collisions of particles with massive moving
scatterers has been considered by S.Ulam \cite{Ulam} in the context
of the unbounded increase of energy in periodically forced Ha\-mil\-to\-nian
systems. It goes back to the question related to the origin of high
energy cosmic particles \cite{Fermi} and known as Fermi acceleration.
The Fermi--Ulam model has been the first system where invariant
curves, chaotic layers and stable islands have been investigated
(details see in \cite{LLBookEn}). It has been shown that in the case
of a quite smooth perturbation of the boundary the particle velocity
is bounded by invariant curves. Otherwise, the velocity can grow
indefinitely.

For low-dimensional {\it integrable} billiards the problem of Fermi
acceleration has been studied on the example of circle and elliptic
billiards \cite{bMarkaryan1, bMarkaryan2, Levi}. In these papers the
authors come to conclusion that the velocity of the particle ensemble
is bounded by the corresponding invariant curves.
Investigations of {\it chaotic billiards} have been performed for the
Lorentz gas \cite{MyZHTPH, MyPHA}. As predicted, perturbations of
the boundary of such a billiard lead to the appearance the Fermi
acceleration for the particle. In addition, the acceleration is higher in
the case of periodical boundary oscillations than in for their stochastic
perturbations.

In the present paper we study so-called stadium-like billiards
\cite{ZaslKriter} which are defined as a closed domain $Q$ with the
boundary $\partial Q$ consisting of two parallel lines and two focusing
curves (Fig.\ref{BTS}). If parameter $b$ is a sufficiently small then
the billiard is a near-integrable system. In this case its fixed points are
stable. As a result, in the stochastic  (or chaotic) "sea" the stability
regions appear which consist of invariant curves. At the same time,
owing to a weak nonlinearity, dynamics near the separatrixes divided
the stability regions of elliptic points is chaotic, and the particle can
reach neighbourhoods of all points in the chaotic layer. In the case of
the fixed boundary the particle dynamics can be both chaotic and
regular, depending on the initial conditions. Introduction of external
perturbations leads to the possibility of the particle passage from
chaotic region to the regular one and back. This is the reason of new
interesting effects which are also described in the paper.

\section{Definitions and maps}

In this section, basic analytical results are presented. They are
necessary for the further description of the billiard dynamics.

\subsection{Stadium-like billiard with the fixed parabolic focusing
com\-po\-nents}

Consider a billiard shown in Fig.\ref{BTS}. To describe its dynamics
let us construct the corresponding dynamical system for $b\ll a$.
For this, we use the known method of specular reflections. It consists
of the reflection of the billiard table form neutral components.
As a result, the stadium is replaced by a caterpillar billiard.
It can be shown that the change in the particle velocity in both cases
is the same. In addition, one can show that between trajectories of the
initial billiard and the obtained "caterpillar", a one-to-one
correspondence takes place.

\begin{figure}[h]
\centering
\includegraphics[width=90mm]{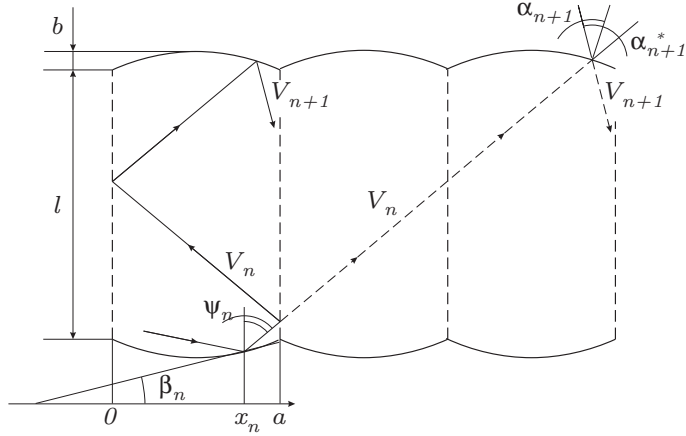}
\caption{\sl {\small A stadium-like billiard and its development.}}
\label{BTS}
\end{figure}

Suppose that at the initial time the particle belongs to the billiard
boundary and its velocity vector directs towards the interior of the
billiard region. Let us choose coordinates $\psi$ and $x$ as shown in
Fig.\ref{BTS}. The motion of the billiard particle induces a map
$(\psi_n,x_n) \to (\psi_{n+1},x_{n+1})$. Suppose that $b\ll l$. In this
case the focusing components can be approximated by the function
$\chi(x)=4bx(x-a)/a^2$. For such billiard configuration the map is
written as follows:
$$
\begin{array}{l}
x_{n+1}=x_{n}+l\tan \psi _{n+1}\ ,\ \mathrm{mod}\ a \ ,\\ \\
\psi _{n+1}=\psi _{n}-2\beta (x_{n+1})\ ,
\end{array}
$$
where $\beta (x)=\arctan\left(\chi^{\prime }(x)\right)$ (see
Fig.\ref{BTS}). If $b$ is small then $\beta \approx 4b(2x-a)/a^{2}$.
For the further analysis change the variables:
$\xi =x/a,$ $\xi \in \lbrack 0,1)$. Then
\begin{equation}
\begin{array}{l}
\xi _{n+1}=\xi _{n}+\displaystyle\frac{l}{a}\tan
\psi _{n}\ ,\ \mathrm{mod}\ 1 \ ,\\ \\
\psi _{n+1}=\psi _{n}-\displaystyle\frac{8b}{a}(2\xi _{n+1}-1)\ .
\end{array}
\label{eqParabolSimple}
\end{equation}

In Fig.\ref{pParabPhP} the phase portrait generated by the map
(\ref{eqParabolSimple}) is shown. Initial conditions for each trajectory
are marked by crosses. One can see that the trajectory with initial
conditions in the chaotic region can reach any point of this region.
At the same time, in the regular regions the points moves along
invariant curves.

\begin{figure}[h]
\includegraphics[angle=-90,width=140mm]{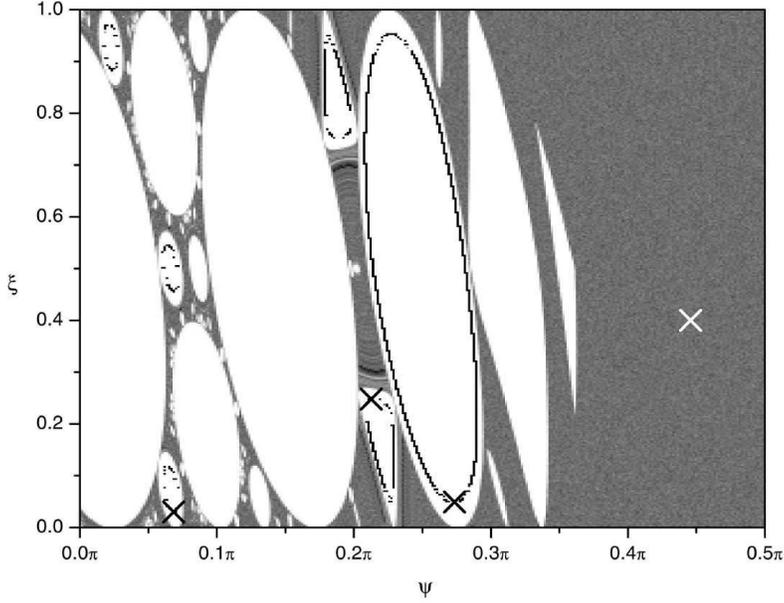} 
\caption{\sl {\small Phase portrait of the billiard with parabolic focusing
components (see map (\ref{eqParabolSimple})) at $a=0.5$,
$b=0.01$ and $l=1$. The diagram contains three regular trajectories
(each by $10^{7}$ iterations) and one chaotic trajectory
($5\cdot 10^{8}$ iterations).}}
\label{pParabPhP}
\end{figure}

It is obvious that the fixed points of the map (\ref{eqParabolSimple})
are the following: $\xi =1/2$ and $\psi _{s}=\arctan (ma/l)$. If $m=0$
then the billiard particle moves strictly vertically. If $m=1$ then it
moves for one cell. And so on. In Fig.\ref{pParabPhP} the fixed points
correspond to maximal ellipses. Let us find the stability conditions
of these points. To this end, change the variables:
$\xi _{n}=\Delta \xi _{n}+1/2$, $\psi _{n}=\Delta \psi_{n}+\arctan
(ma/l)$ and linearize the map. Then we get:
$$
\begin{array}{l}
\Delta \xi _{n+1}=\Delta \xi _{n}+
\displaystyle\frac{l}{a\cos ^{2}\psi _{s}}\Delta
\psi_{n}\ +O(\Delta \psi _{n}^{2})\ , \\ \\
\Delta \psi _{n+1}=\Delta \psi _{n}-
\displaystyle\frac{16b}{a}\Delta \xi_{n+1}\ ,
\end{array}
$$
where $\psi _{s}=\arctan (ma/l)$. The corresponding
transformation matrix has the form:
$$
A=\left(
\begin{array}{cc}
1 & \displaystyle\frac{l}{a\cos ^{2}\psi _{s}} \\
-\displaystyle\frac{16b}{a} & 1-
\displaystyle\frac{16bl}{a^{2}\cos ^{2}\psi _{s}}
\end{array}
\right)\ .
$$
It is not hard to see that $\det A=1$. Thus, the map preserves the
phase volume.

The stability criterion for the fixed points is
$\left|{\rm Tr A}\right|\leq 2$. Then $\cos ^{2}\psi _{s}\geq 4bl/a^{2}$
or $m^{2}\leq l/(4b)- l^{2}/a^{2}$. On the other hand, transition
to chaos take place if
\begin{equation}
\frac{{4bl}}{a^{2}}>1\ .
\label{eStadSimpleStochCond}
\end{equation}
Eigenvalues of the matrix $A$ are $\lambda _{1,2}=e^{\pm i\sigma}$,
where $\cos\sigma=\displaystyle\frac{1}{2}{\rm Tr} A$. Let us
introduce $f=l/(a\cos ^{2}\psi _{s})$, $g=16b/a$. In this case
$$
A=\left( \begin{array}{cc}
1 & f \\ -g & 1-fg
\end{array}
\right)
$$
and its eigenvalues
$$
X_{1,2}=\left( \begin{array}{c}
1 \\ \displaystyle\frac{e^{\pm i\sigma }-1}{f}
\end{array}
\right)\ .
$$
Consider matrix $X$ with the columns of eigenvectors. As known,
in this case the matrix $\Lambda =X^{-1}AX$ is diagonalization of
the matrix $A$:
$$
\Lambda =\left( \begin{array}{cc}
e^{i\sigma } & 0 \\ 0 & e^{-i\sigma }
\end{array}
\right)\ .
$$
New variables for which the transformation matrix has a diagonal
form are the following:
$$
\left( \begin{array}{c}
Z \\ Z^{\ast }
\end{array}
\right) =X^{-1}\left( \begin{array}{c}
\Delta \xi \\ \Delta \psi
\end{array}
\right)\ ,
$$
where
$$
X^{-1}=\frac{i}{2\sin \sigma }\left( \begin{array}{cc}
e^{-i\sigma }-1 & -f \\ -e^{i\sigma }+1 & f
\end{array}
\right)\ .
$$
One can see that $Z$ and $Z^{\ast}$ are complex conjugate. Thus,
$$
Z_{n+1}=Z_{n}e^{i\sigma }\ .
$$
If we take $Z$ in the form of $Z=Ie^{i\theta }$ then
in the action--angle variables we obtain:
$$
\begin{array}{cc}
I_{n+1} = I_{n} \ ,\\
\theta _{n+1} = \theta _{n}+\sigma \ .
\end{array}
\label{eqRotMap}
$$
Thus, the particle motion around the stable point is written
by the map with the following rotation number:
\begin{equation}
\sigma = \arccos \left( 1-\frac{8bl}{a^{2}\cos ^{2}\psi _{s}}\right) \ .
\label{eqSigma}
\end{equation}
In turn, old variables have the form:
$$
\begin{array}{l}
\Delta \xi =2I\cos \theta \ , \\
\\
\Delta \psi =\displaystyle\frac{2I}{f}\left(\cos (\sigma +
\theta )-\cos \theta \right)\ .
\end{array}
\label{eqBackMap}
$$
Therefore, the Jacobian of the transformation is
$J=\displaystyle-\frac{4I}{f}\sin\sigma$.

\subsection{Perturbations of the boundary and resonance}

The particle motion with the velocity $V$ generates a flow for which
we can introduce time $t$. In turn, the time between two sequential
collisions is the following:
$\tau \approx \displaystyle\frac{l}{\cos \psi _{s}}\frac{1}{V}$.
Thus, we get the rotation period:
$$
T_{1}=\frac{2\pi }{\sigma }\tau =\frac{2\pi l}{\cos \psi _{s}\arccos \Bigl(
1-8bl/(a\cos\psi )^{2}\Bigr) V}\ .
\label{eqPeriod}
$$

If the system undergoes external perturbations of the boundary
of period $T_{ext}$ then for $T_{1}=T_{ext}$ we can observe the
resonance between degrees of freedom. This leads to the fact that
regions including stability areas (see Fig.\ref{pParabPhP}) are
accessible for the particle.

Some words about the nature of the resonance (see
Fig.\ref{pInvarCurves}). At the motion along the invariant curve
in the neighbourhood of the stable point, angle $\psi$ oscillate.
Collision of the particle with the perturbed boundary leads to the
change in $\psi$. If the boundary moves towards the particle, then
the angle decreases. Otherwise it increases. Suppose that
the image of the trajectory moves along the arc $AB$. In this case,
if the particle undergoes collisions coming from the opposite
side, then the trajectory tends to inside of the area.
If the boundary and the particle velocities have the same direction,
then the trajectory goes outside of the region.

For the resonance between the boundary oscillations and the motion
along the invariant curve form $B$ to $A$, collisions change the
particle direction. This leads to larger shift of the trajectory to the
stable point.

\begin{figure}[h]
\centering
\includegraphics[width=90mm]{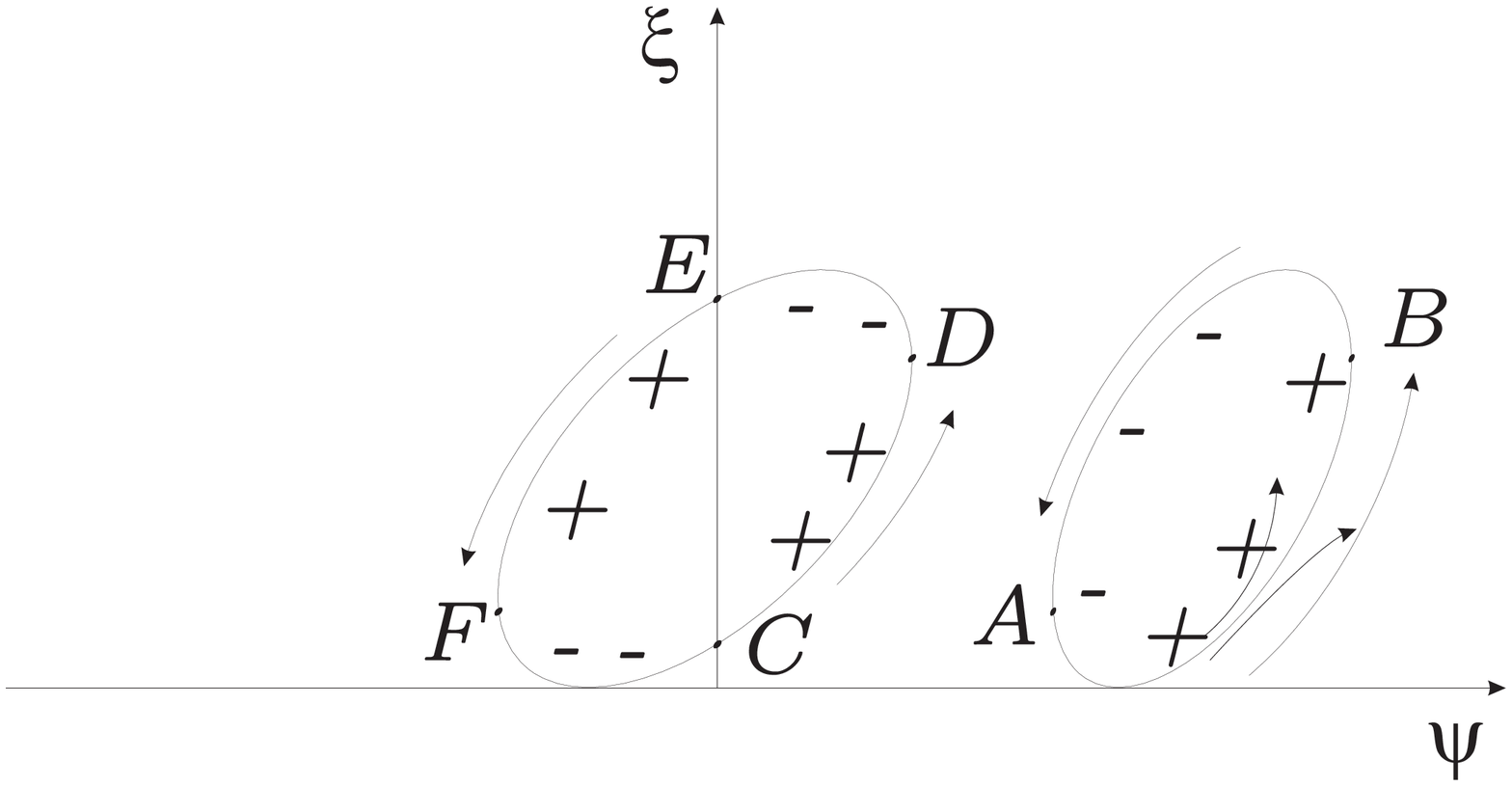}
\caption{\sl {\small Invariant curves around stable points
(schematically).}}
\label{pInvarCurves}
\end{figure}

From the equality $T_{1}=T_{ext}$ we can obtain the resonance
condition for the particle velocity:
\begin{equation}
V_{r}=\frac{l}{\cos \psi _{s}\arccos\Bigl(1-
8bl/(a\cos \psi _{s})^{2}\Bigr)}\ .
\label{eqVResonans}
\end{equation}
Hereafter, for simplicity we assume that the frequency of external
perturbations $\omega =1$. Then $T_{ext}=2\pi $.

For the invariant curves with $\psi _{s}=\arctan (ma/l)$, $m\geq 1$,
there exists only one area where the absolute value of the angle
$\psi _{n}$ increases (along the arc $AB$ from $A$ to $B$ in
Fig.\ref{pInvarCurves}), and the area of the decrease in $\psi _{n}$.
However, for the central stable fixed point there are two such areas:
$CD$, $EF$ for increasing and $DE$, $FC$ for decreasing. This
leads to the fact that in the neighbourhood of this point resonance is
observed for the same condition that in the other areas, but the
particle velocity is less by half. Therefore,
\begin{equation}
V_{r}^{0}=\frac{V_{r}}{2}=\frac{l}{\arccos\left(1-8bl/a^{2}\right)}\ .
\label{eqVResCentr}
\end{equation}

\subsection{Focusing components in the form of the circle arcs}

In this section, we consider a stadium-like billiard with the boundary
consisting of two focusing components in the form of the circle arcs
and compare the obtained results with the previous parabolic case.

\subsubsection{Fixed boundary}

Suppose that focusing components are arcs of the radius $R$ circle
(symmetrical about the vertical billiard axis) and the angle measure
$2\Phi$ (Fig.\ref{pCirBill}). Geometrically we can obtain that
$$
R=\frac{a^{2}+4b^{2}}{8b}\ ;\ \Phi =\arcsin \frac{a}{2R}\ .
$$
For such a billiard the chaoticity condition is obtained as follows
\cite{BnchCondStoch}. Assume that $Q\subset R^{2}$ and
the focusing component is a part of the circle $C$. Chaos
can be observed if disk $D$, $\partial D=C$, belongs to the
billiard table $Q$. Thus,
$$
\frac{l}{2R}=\frac{4bl}{a^{2}}>1\ ,
$$
that is the same as (\ref{eStadSimpleStochCond}) obtained from
the analysis of the stable points.

\begin{figure}[h]
\centering
\includegraphics[width=90mm]{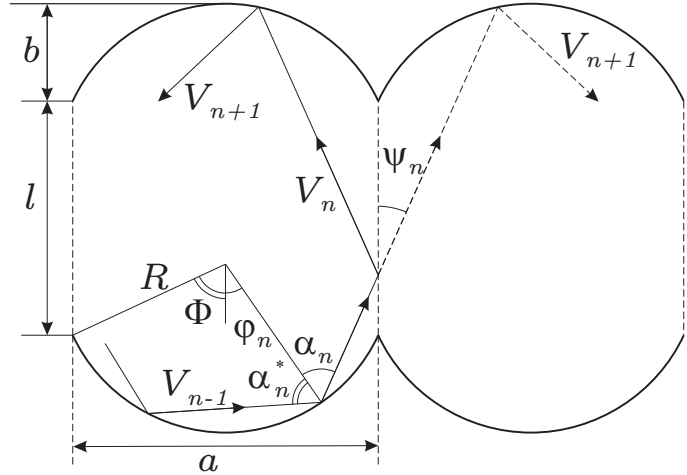}
\caption{\sl {\small A stadium-like billiard with focusing components
in the form of circle arcs.}}
\label{pCirBill}
\end{figure}

Let us introduce dynamical variables as shown in Fig.\ref{pCirBill}.
Assume that angles $\varphi _{n}$ and $\alpha _{n}^{\ast }$
are counted counterclockwise, and the angle $\alpha _{n}$
is counted clockwise. For the fixed boundary
$\alpha _{n}^{\ast}=\alpha _{n}$. Suppose that $V_{n}$
is the particle velocity, and $t_{n}$ is a time of $n$-th collision.
Let us find a map describing dynamics of the particle in such
a billiard. For this it is necessary to consider two cases:
1) After collision with the focusing component the particle
collides with the same component (multiple collisions);
2) After the collision, the particle moves to the opposite
focusing component.

\begin{figure}[h!]
\centering
\includegraphics[width=90mm]{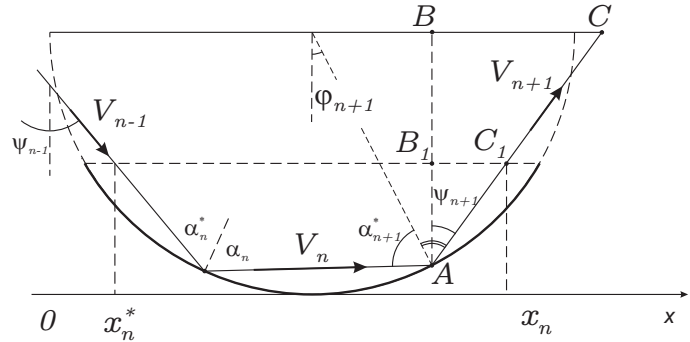}
\caption{\sl {\small Multiple collisions with the focusing component.}}
\label{pCirComp}
\end{figure}

1) Multiple collisions (Fig.\ref{pCirComp}).

In this case, geometrically we get the following map:
\begin{equation}
\begin{array}{l}
\alpha _{n+1}^{\ast }=\alpha _{n}\ , \\ \\
\alpha _{n+1}=\alpha _{n+1}^{\ast }\ , \\ \\
\varphi _{n+1}=\varphi _{n}+\pi -2\alpha _{n}\ (\mathrm{mod\ }2\pi )\ , \\
\\ t_{n+1}=t_{n}+\displaystyle\frac{2R\cos \alpha _{n}}{V_{n}}\ .
\end{array}
\label{eqUnPertCirBill1}
\end{equation}
If $|\varphi_{n+1}|<\Phi$, then the particle collides with the same
component. Otherwise, $n+1$-th collision with the opposite
components takes place.

2) Collision with opposite components.

For this case the map can be written as follows:
\begin{equation}
\begin{array}{l}
\alpha _{n+1}^{\ast }=\arcsin \left[ \sin \left( \psi _{n}+\Phi \right)
-\displaystyle\frac{x_{n+1}^{\ast }}{R}\cos \psi _{n}\right] \ , \\ \\
\alpha _{n+1}=\alpha _{n+1}^{\ast }\ , \\ \\
\varphi _{n+1}=\psi _{n}-\alpha _{n+1}^{\ast }\ , \\ \\
t_{n+1}=t_{n}+\displaystyle\frac{R(\cos \varphi _{n}+
\cos \varphi _{n+1}-2\cos \Phi)+l}{V_{n}\cos \psi _{n}}\ ,
\end{array}
\label{eqUnPertCirBill2}
\end{equation}
where
$$
\begin{array}{ccc}
\psi _{n} &=&\alpha _{n}-\varphi _{n}\ , \\
x_{n} &=&\displaystyle\frac{R}{\cos \psi _{n}}
\left[ \sin \alpha _{n}+\sin \left( \Phi - \psi _{n}\right) \right] \ , \\
x_{n+1}^{\ast } &=&x_{n}+l\tan \psi _{n}\ (\mathrm{mod\ }a)\ .
\end{array}
$$
Really, let us extend the circle arc up to the semicircle
(Fig.\ref{pCirComp}). Introduce the angle $\psi $ between
the vertical and the velocity vector. It is counted clockwise.
It is obvious that $BC=R\cos\varphi _{n+1}\tan \psi _{n+1}$.
From triangles $ABC$ and $AB_{1}C_{1}$ we get:
$\displaystyle\frac{B_{1}C_{1}}{BC}=\frac{AB_{1}}{AB}=
\frac{R(\cos\varphi _{n+1}-\cos \Phi )}{R\cos \varphi _{n+1}}$.
Therefore,
$$
B_{1}C_{1}=R\tan \psi _{n+1}(\cos \varphi _{n+1}-\cos \Phi )\ .
$$
Now, using the value $x_{n}$ for the next collision we obtain:
\begin{equation}
\begin{array}{l}
x_{n+1}=R\sin \Phi +R\sin \varphi _{n+1}+
R\tan \psi _{n+1}(\cos \varphi_{n+1}-\cos \Phi )= \\
\\
=\displaystyle\frac{R}{\cos \psi _{n}}\left( \sin \alpha _{n}+
\sin (\Phi -\psi_{n})\right) \ .
\end{array}
\label{xn}
\end{equation}
In addition,
$$
x_{n+1}^{\ast }=x_{n}+l\tan \psi _{n}\ (\mathrm{mod\ }a)\ .
$$
If we invert the particle motion then the expression (\ref{xn})
gives connection between $x_{n}^{\ast }$ and $\alpha _{n}^{\ast }$.
This fact allows us to get the explicit system (\ref{eqUnPertCirBill2}).

\subsubsection{Perturbed boundary}

Suppose that focusing components are perturbed in such a way that
their velocity in each point is the same and directed by the normal to
the component. Assume that the velocity value depends on time
as follows: $U(t)=U_{0}f(\omega (t+t_{0}))$, where $\omega$ is
a frequency oscillation. We consider the case $U_{0}/\omega\ll l$,
i.e. the shift of the boundary is small enough and can be neglected.
Therefore, the billiard map is written as follows:
\begin{equation}
\begin{array}{l}
V_{n}=\sqrt{V_{n-1}^{2}+4V_{n-1}\cos \alpha _{n}^{\ast }U_{n}+4U_{n}^{2}}\ ,
\\ \alpha _{n}=\arcsin \left(\displaystyle\frac{V_{n-1}}{V_{n}}
\sin \alpha _{n}^{\ast}\right) \ ,
\end{array}
\label{eqPertCirBill2} 
\end{equation}
\begin{equation}
\left.
\begin{array}{l}
\alpha _{n+1}^{\ast }=\alpha _{n}\ , \\
\varphi _{n+1}=\varphi _{n}+\pi -2\alpha _{n}\ (\mathrm{mod\ }2\pi )\ , \\
t_{n+1}=t_{n}+\displaystyle\frac{2R\cos \alpha _{n}}{V_{n}}\ ,
\end{array}
\right\} \qquad if\ |\varphi _{n+1}|\leq \Phi
\label{eqPertCirBill3} 
\end{equation}
\begin{equation}
\left.
\begin{array}{l}
\psi _{n}=\alpha _{n}-\varphi _{n}\ , \\
x_{n}=\displaystyle\frac{R}{\cos \psi _{n}}\left[\sin\alpha _{n}+
\sin \left( \Phi -\psi_{n}\right) \right] \ , \\
x_{n+1}^{\ast }=x_{n}+l\tan \psi _{n}\ (\mathrm{mod\ }a)\ , \\
\alpha _{n+1}^{\ast }=\arcsin \left[ \sin \left( \psi _{n}+\Phi \right)
-\displaystyle \frac{x_{n+1}^{\ast}}{R}\cos \psi _{n}\right] \ , \\
\varphi _{n+1}=\psi _{n}-\alpha _{n+1}^{\ast }\ , \\
t_{n+1}=t_{n}+\displaystyle\frac{R(\cos \varphi _{n}+
\cos \varphi _{n+1}-2\cos \Phi)+l}{V_{n}\cos \psi _{n}}\ .
\end{array}
\right\} \qquad if\ |\varphi _{n}+\pi -2\alpha _{n}|>\Phi
\label{eqPertCirBill4}
\end{equation}
The given map describes a stadium-like billiard with the focusing
components in the form of the circle arcs. This map is exact one
except for the approximation $U_{0}/\omega\ll l$. The first group
(\ref{eqPertCirBill3}) corresponds to sequential multiple collisions
with one of the focusing components, and the second group
(\ref{eqPertCirBill4}) corresponds to the passage to the opposite
side of the boundary.

\section{Numerical analysis}

In this Section we consider stadium-like billiards with constant and
perturbed bo\-un\-da\-ries. In the first case, the particle dynamics is
described by the exact map (\ref{eqParabolSimple}) and approximate
map (\ref{eqUnPertCirBill1})--(\ref{eqUnPertCirBill2}), respectively.

\subsection{Phase diagrams of billiards with the fixed boundaries}

Phase portrait can help to understand the system dynamics
and define chaotic and regular regions in its phase space.
In Fig.\ref{pParabPhP} the phase diagram of the
billiard with the fixed parabolic boundary is shown (see map
(\ref{eqParabolSimple}). Crosses in this figure are initial
conditions. One can see that phase plane is divided into
regular and chaotic regions. If the initial conditions belong
to the regular region then the trajectory remains here and
forms the corresponding invariant curves. However, beginning in
chaotic region the phase trajectory uniformly covers this region.
The described portrait has been obtained on the basis of
three regular trajectories (each contains $10^{7}$ iterations) and
one chaotic trajectory ($5\cdot 10^{8}$ iterations) of the map
(\ref{eqParabolSimple}). Geometric size of the billiard is the
following: $a=0.5,$ $b=0.01,$ $l=1$.

\begin{figure}[h]
\caption{\sl {\small Phase portrait of stadium-like billiard with
focusing components in the form of circle arcs (see map
(\ref{eqUnPertCirBill1})--(\ref{eqUnPertCirBill2})).
Parameters of the billiard are the same as in Fig.\ref{pParabPhP}.
One can see nonuniformity of the covering of the chaotic
region.}}
\label{pCirStablePhP}
\includegraphics[angle=-90,width=130mm]{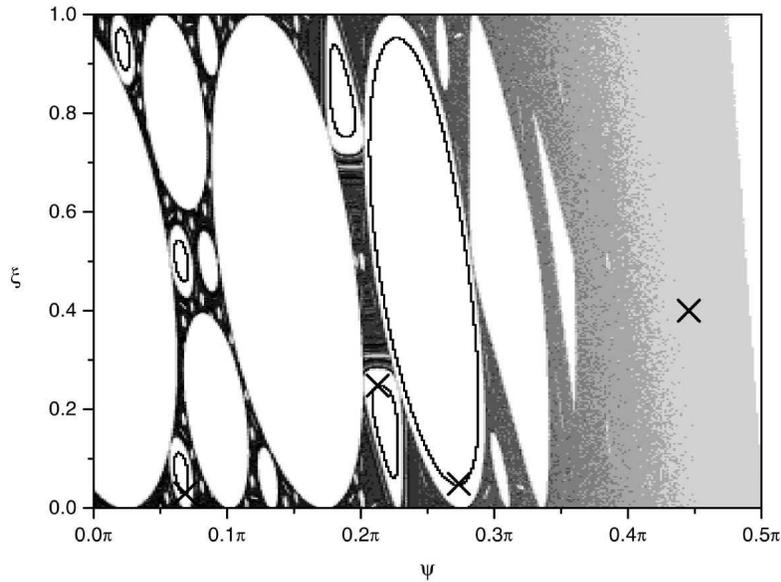}
\centering
\end{figure}

In Fig.\ref{pCirStablePhP} similar results concerning the map
(\ref{eqUnPertCirBill1})--(\ref{eqUnPertCirBill2}), where the depth
of the focusing components is taken into account, are shown.
Remind that for this map the approximation $b\ll l,a$ has been
used. To reasonable comparison, the geometric sizes of the billiard
and the number of trajectories have been chosen the same as
in the previous case (Fig.\ref{pParabPhP}).

The difference between obtained diagrams can be easily explain
by means of Fig.\ref{pDiffer}.

\begin{figure}[h]
\centering
\includegraphics[width=90mm]{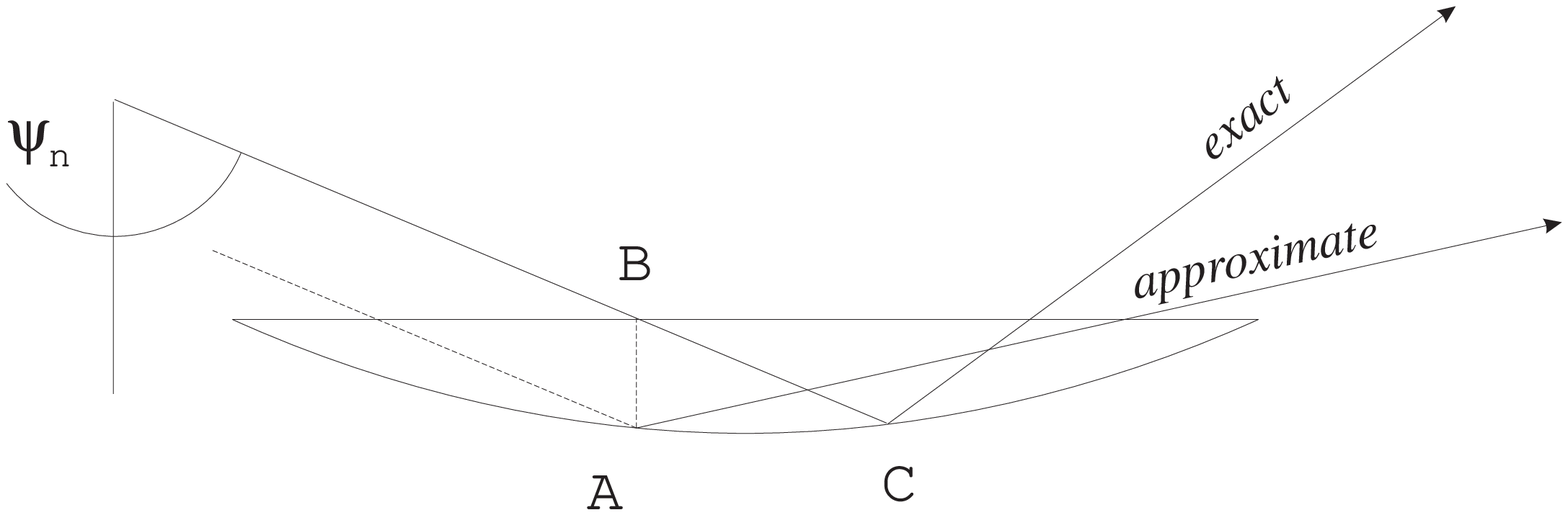}
\caption{\sl {\small The difference between the exact billiard map and
approximate map.}}
\label{pDiffer}
\end{figure}

Approximation of a small enough depth of the focusing component
for the map (\ref{eqParabolSimple}) means the following. Let $B$
be an intersection point of the parabola ends and the particle
trajectory. In approximation, we consider the particle collision in
the point $A$ which is a projection of $B$ into the arc. But, in fact,
the point $C$ is the collision point. As a result, in the exact case for
large enough $\psi$, collisions happen mainly with the right (in the
Figure) part of the arc, and from collision to collision the angle $\psi$
decreases. Thus, the billiard particles as if "push out" to the region
of the small angel $\psi$. This is in agreement with
Fig.\ref{pCirStablePhP} where the region $\psi<\pi/2$ is empty..

\subsection{Perturbed map}

In this section we consider the problem of the velocity change
depending on its relation to the resonance value (\ref{eqVResonans}).

\subsubsection{Phase diagrams}

Consider the map (\ref{eqPertCirBill2})--(\ref{eqPertCirBill4})
corresponding to the perturbed stadium-like billiards. Construction
of the phase diagrams have been performed for the same values of
geometric parameters as in the previous \S3.1. Therewith, the
amplitude of os\-cil\-la\-ti\-ons $U_{0}=0.01$. As noted above (see \S2.2),
for various particle velocities the corresponding phase portraits should
be different form each other.

\begin{figure}[h]
\centering
\includegraphics[angle=-90,width=100mm]{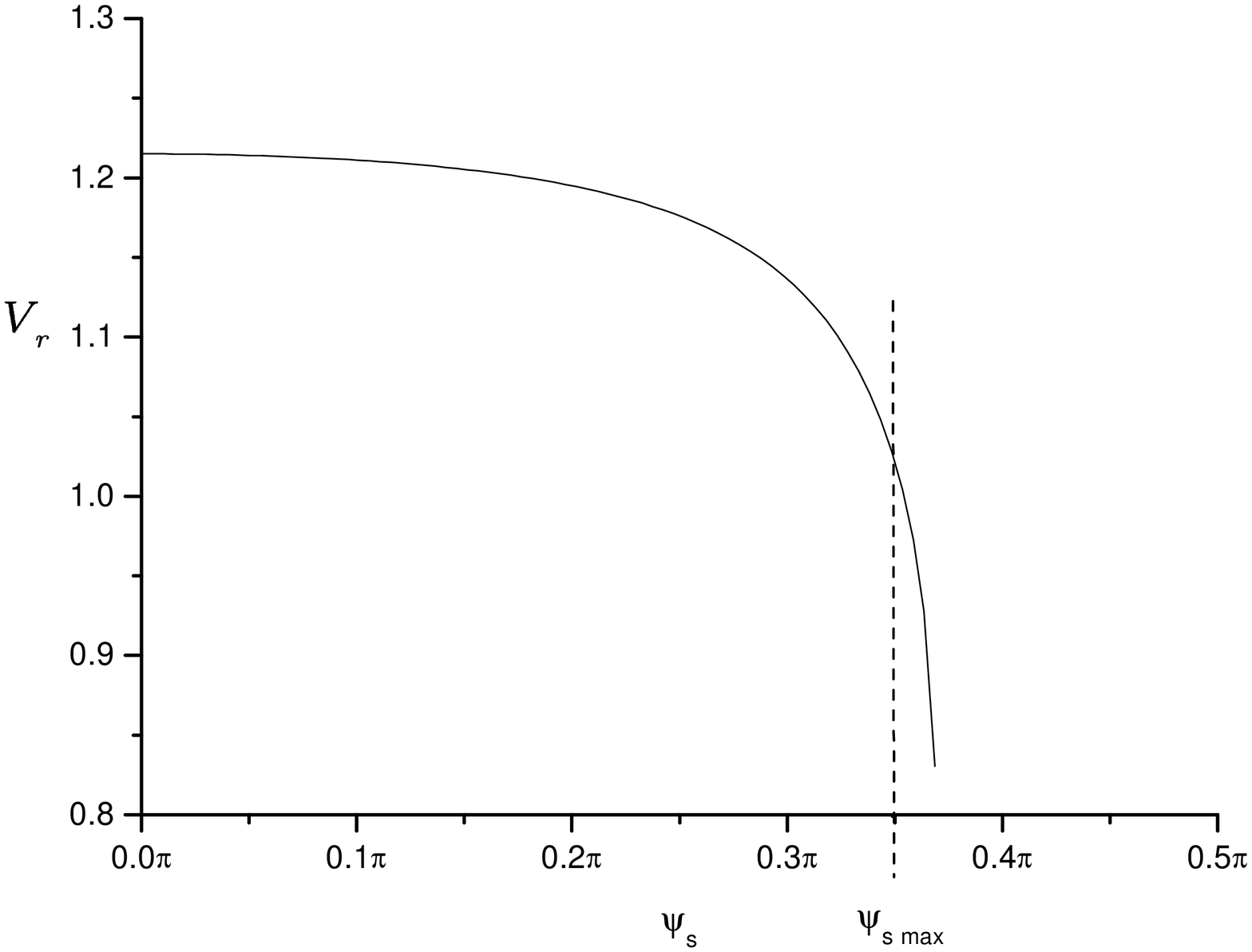}
\caption{\sl {\small The resonance velocity as a function of
$\psi _{s}$ (see (\ref{eqVResonans})).}}
\label{pVCrit}
\end{figure}

In Fig.\ref{pVCrit} the resonance velocity as a function of the angle
$\psi _{s}$ (see (\ref{eqVResonans})) is shown. One can see that in
the region form $0$ to $\psi _{s\ \max}$ (where $\psi _{s\ \max}$ is a
maximal angle for which the fixed points are still stable) the value of
the resonance velocity is varied through a small range.

Phase portraits for the perturbed billiard are shown in
Fig.\ref{pCirPertPhP}. For detailed nu\-me\-ri\-cal analysis
three particle ensembles with initial value $V_{0}=1$, $1.2$
and $1.5$ have been considered. Therewith initial conditions
have been chosen in the chaotic region in a random way.
Thus, the obtained portraits give an insight into the velocity
change of the billiard particle. In the obtained Figure, the vertical
shaded areas correspond to the velocity increasing, and the horizontal
shaded areas fit its decreasing. The wait areas (without shading) are
the intermediate ones; here the particle velocity is transient. The black
tones correspond to the areas which are inaccessible for the phase
trajectory.

\begin{figure}[p]
\centering
\includegraphics[height=200mm]{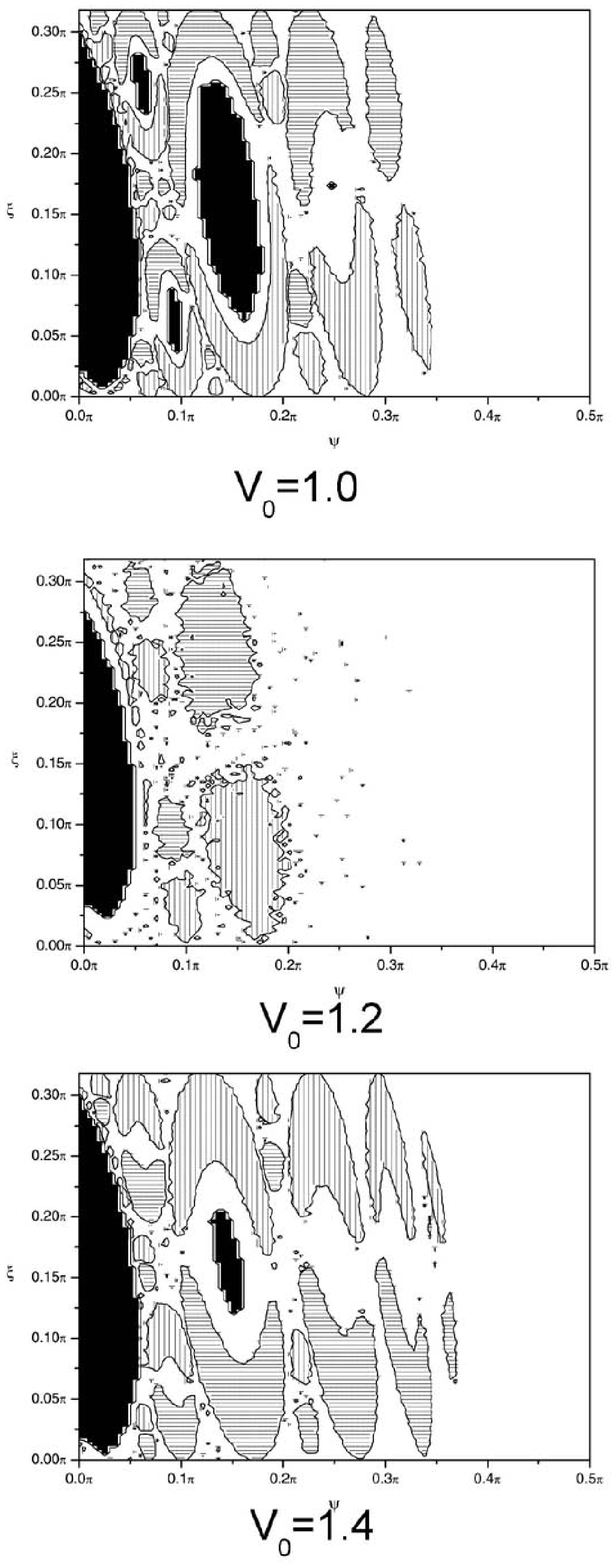}
\caption{\sl {\small Phase diagrams of the velocity change in the
billiard with the perturbed boundary (see
(\ref{eqPertCirBill2})--(\ref{eqPertCirBill4})) at
$b=0.01$, $a=0.5$, $l=1$, $U_{0}=0.01$ and $\protect\omega =1$.
$V_{0}=1.0$, $1.2$ (resonance), $1.5$.}}
\label{pCirPertPhP}
\end{figure}

\begin{figure}[p]
\centering
\includegraphics[height=200mm]{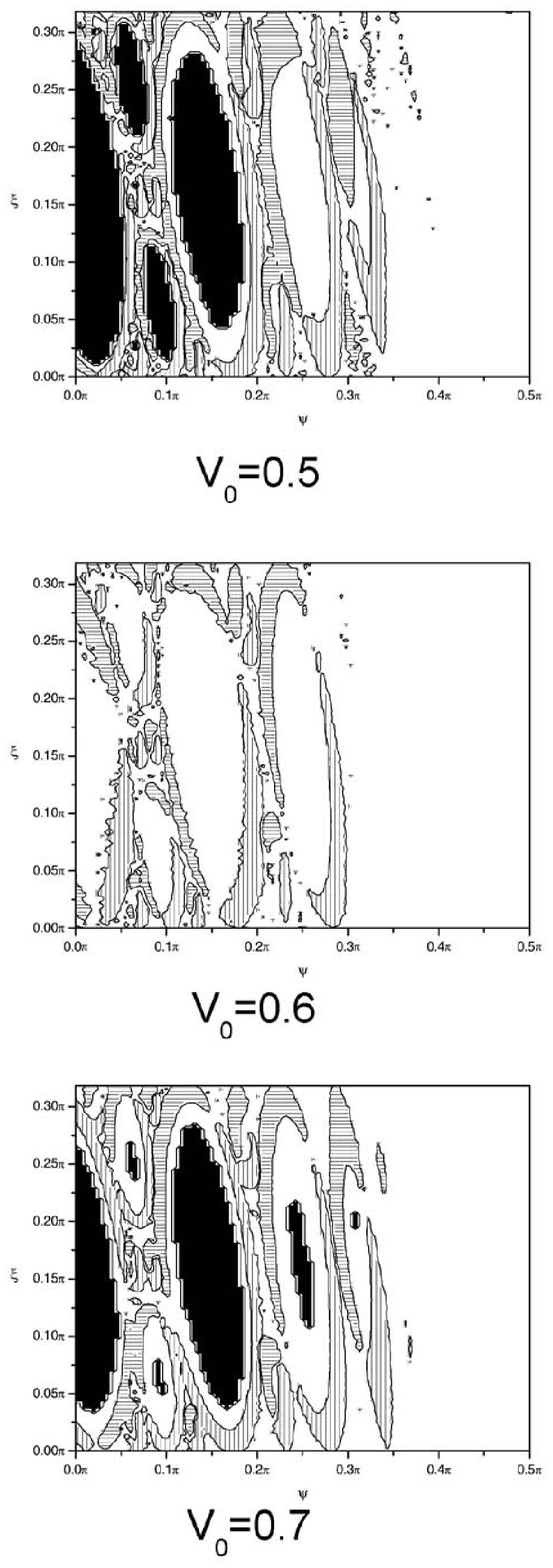}
\caption{\sl {\small The same as in Fig.\ref{pCirPertPhP} but
$V_{0}=0.5$, $0.6$ and $0.7$.}}
\label{pCirPertPhP2}
\end{figure}

As follows from obtained diagrams, if $V_{0}\neq V_{r}$ then
around the stable fixed points there exist the areas surrounded by
invariant curves. As before, these areas are inaccessible for
particles from the chaotic regions. At the same time, in the
neighbourhood which has become accessible for the particles
as a result of per\-tur\-ba\-tions, one can see areas of the increasing
and decreasing velocity. Depending on the relation to the resonance
velocity value, they can change places.

If $V_{0}=V_{r}$ (the resonance harmonic) then all neighbourhoods of
the stable fixed points (except for the central one, $\psi _{0}=0$,
$\xi _{0}=1/2$) become accessible for the trajectory. In addition, for
this resonance there are no the well-defined areas where the
particles have an acceleration.

Following (\ref{eqVResCentr}) the resonance velocity in
the neighbourhood of the central stable fixed points
$V_{r}^{0}\approx 0.6$ (primary resonance). In Fig.\ref{pCirPertPhP2}
the phase diagrams for the initial velocities $V_{0}=1$, $1.2$ and
$1.5$ are shown. One can see that at $V_{0}=V_{r}^{0}$ all areas
of the phase space are accessible.

\subsubsection{The particle velocity as a function of iterations}

Numerical investigations of the perturbed billiard described by
the map (\ref{eqPertCirBill2})--(\ref{eqPertCirBill4}) have been
performed in two cases: when the billiard has strong chaotic
properties and for a near-rectangle stadium. In the first case
the billiard is a "classical" stadium. Then $\Psi=\pi/2$
and the billiard is a domain with a boundary that consists of two
semicircles and two parallel segments tangent to them. The latter
case means that focusing components are segments of the almost
straight line, and the billiard system is a near-integrable one.

For the first case the following billiard parameters were chosen:
$a=0.5$, $b=0.25$, $l=1$, $u_0=0.01$, $\omega=1$, and
$V_0=0.1$. The particle velocity was cal\-cu\-la\-ted as the
averaged value of the ensemble of 5000 trajectories with different
initial conditions (solid curve 1 in Fig.\ref{pVn}). These initial
conditions were different from each other by a random choice of the
direction of the velocity vector $v_0$. As follows from the numerical
analysis, the obtained dependence has approximately the
square-root behaviour $\left(V(n)\sim\sqrt n\right)$.
The fitting function $y\sim an^c$
(the dot-and-dash curve 1 in Fig.\ref{pVn}) yields the following
values: $a=0.01015\pm 0.00002$ and $c=0.4446\pm 0.0002$.

A near-integrable case means that parameter $b$ (see
Fig.\ref{pCirBill}) is a sufficiently small, and the curvature of the
focusing components gives rise only weak nonlinearity in the system.
In such a configuration the billiard phase space has regions with
regular and chaotic dynamics. This case is much more interesting for
investigations.

As follows from numerical investigations, on each side
of the resonance the behaviour of the particle velocity is essentially
different. If the initial value $V_0<V_c$ then the particle velocity
decreases up to a certain quantity $V_{fin}<V_c$, and the particle
distribution tends to the stationary one in the interval $(0,V_{fin})$. If,
however, $V_0>V_c$ then billiard particles can reach high velocities.
In this case the particle distribution is {\it not} stationary, and it grows
infinitely. In addition, the average particle velocity is also not bounded.

For detailed numerical investigations initial conditions were
randomly chosen in the chaotic region of the unperturbed billiard.
In Fig.\ref{pVn} the particle velocity as a function of the number of
iterations is shown (curves 2--5). The billiard parameters remains the
same as for the stadium-like billiard (curve 1) except for $b=0.01$.

\begin{figure}[h]
\centering
\includegraphics[angle=-90,width=120mm]{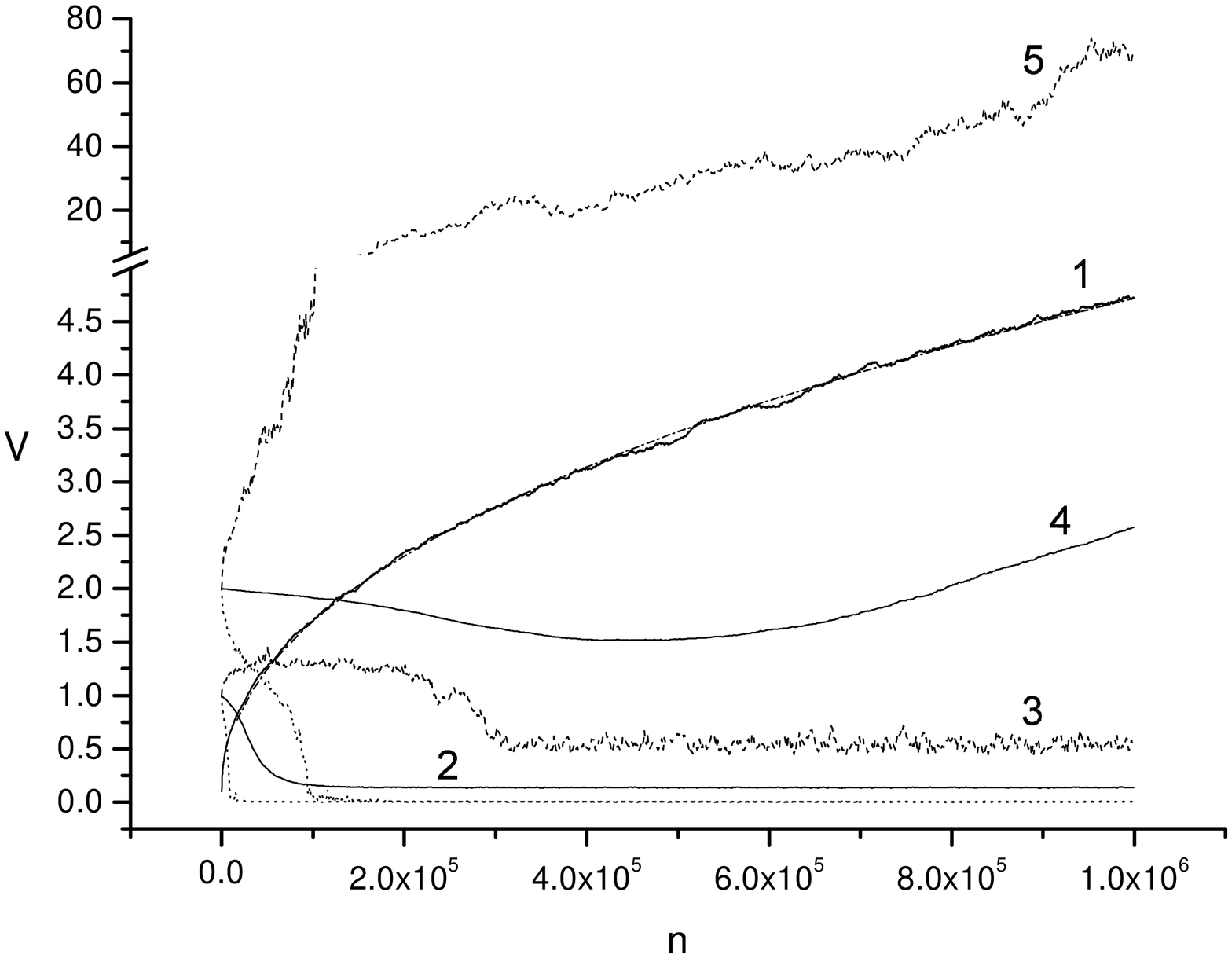}
\caption{\sl {\small Average velocity of the ensemble of 5000 particles
in a stadium as a function of the number of collisions, $l=1$,
$a=0.5$, $U_0=0.01$ and $\omega =1$. Two (dot-dash and solid)
curves 1 corresponds to the billiard with strong chaotic properties
($b=0.25$). Curves 2--5 correspond to the near-integrable system
($b=0.01$): $V_{0}=1$ (curve 2, 3) and $V_{0}=2$ (curve 4, 5).
Curves 2 and 4 are the average velocities of the particle ensemble.
Curves 3 and 5 correspond to maximal velocities reached by the
particle ensemble to the $n$-th iteration.}}
\label{pVn}
\end{figure}

On the basis of 5000 realisations and for every initial velocity,
{\it three curves} have been constructed: the average, minimal and
maximal velocities which the particle ensemble has reached to the
$n$-th iteration. So, we can see the interval of the velocity change.
As follows from this Figure, if $V_0<V_c$ then the averaged particle
velocity (solid curve 2) gradually decreases and tends to a constant.
The maximal velocity of particles (dotted curve 3) also
decreases up to $V_{fin}$ and then fluctuate near this value.
Eventually, the particle velocities lie in the interval $0<V<V_{fin}$.
In the case of $V>V_c$, the minimal velocity of particles as before
decreases. This means that in the ensemble there is a number
of particles which are in the region of low velocity values.
In our numerical analysis the part of such particles was about
75 percents. At the same time, there are particles with high
velocities (dashed curve 5 which corresponds to the maximal
velocity of the ensemble). As a result, the averaged particle
velocity (solid curve 4) increases.

\begin{figure}[h]
\centering
\includegraphics[angle=-90,width=120mm]{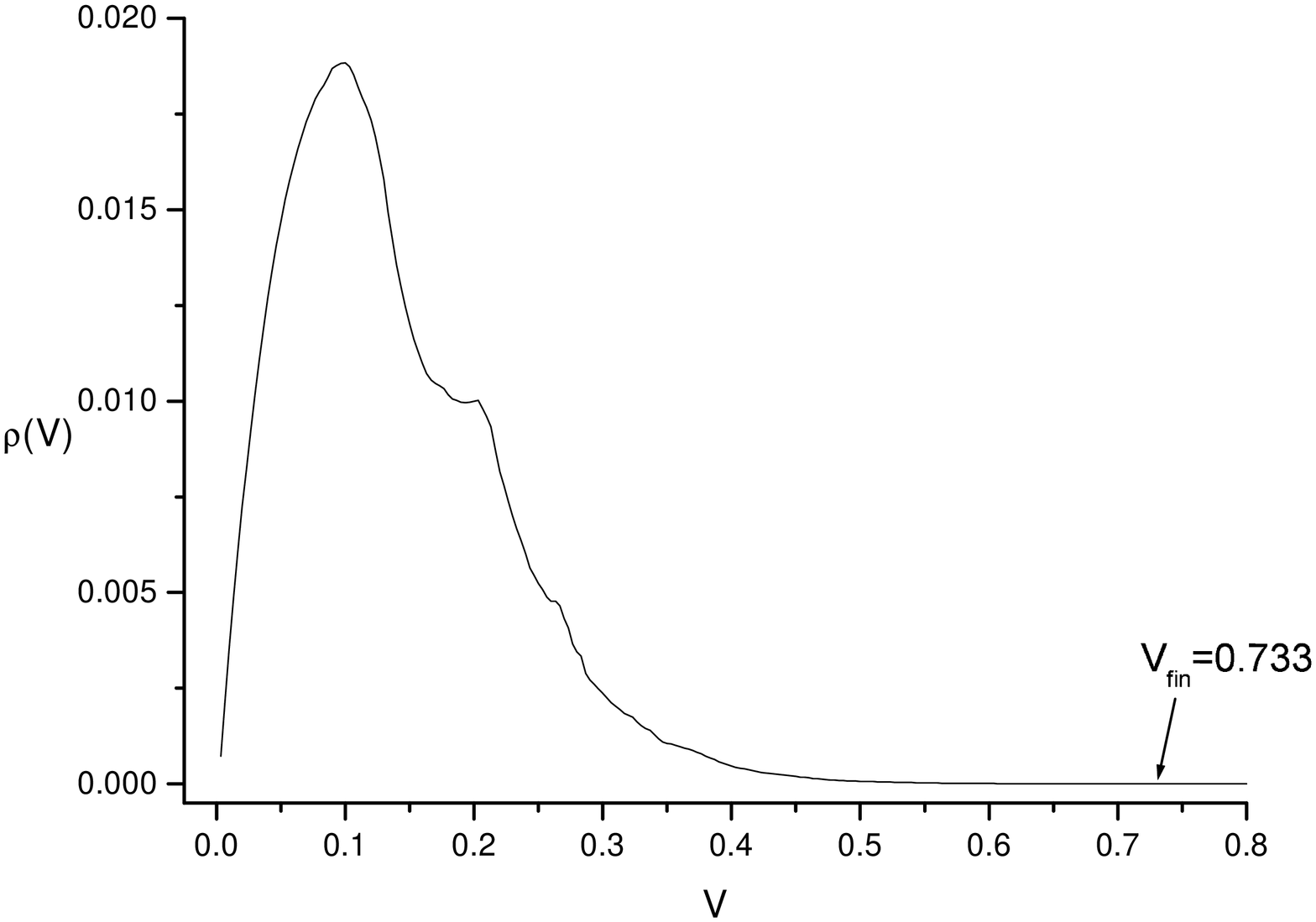}
\caption{\sl {\small Stationary distribution of the particle velocity
calculated by the one particle trajectory during $10^9$ iterations.
$V_{fin}$ is a maximally reached velocity.}}
\label{pRhoV}
\end{figure}

In Fig.\ref{pRhoV}  a stationary velocity distribution is shown. This
distribution was cal\-cu\-la\-ted by the one particle trajectory during $10^9$
iterations. The initial velocity was chosen as follows:
$V_0\approx V_{fin}/2$. The value denoted by $V_{fin}$ corresponds
to the maximally reached velocity.

\section{Concluding remarks}

Billiards are very convenient models of several physical systems.
For example, par\-ti\-cle trajectories in billiards of specific configuration
can be used in modelling a lot of dynamical systems. Moreover, most
approaches to the problems of mixing in many-body systems go back
to billiard-like questions. A natural physical generalisation of a billiard
system is a billiard whose boundary is not fixed, but varies by a certain
law. This is a quite new field which opens new prospects in studies of
problems that have been known for a long time.

In the present article we have studied the problem of the billiard ball
dynamics in a stadium with the periodically perturbed boundary.
Numerical analysis shows that for the
case of the developed chaos, the dependence of the particle
velocity on the number of collisions has the root character.
At the same time, for a near-rectangle
stadium an interesting phenomena is observed. Depending on
the initial values, the particle ensemble can be accelerated,
or its velocity can decrease up to the quite low magnitude.
However, if the initial values do not belong to a chaotic
layer then for quite high velocities the particle acceleration
is not observed.

Analytical description of the considered phenomena
(mechanisms of deceleration and acceleration of the billiard particle,
stabilisation of unstable points etc.) required more detailed
analysis and will be published soon \cite{bTBPubl}.


\baselineskip=16pt

\begin{thebibliography}{99}

\footnotesize

\bibitem{Birkgof}
G.Birkhoff. {\it Dynamical Systems.}--- American
Mathematical Society, N.Y., 1927.

\bibitem{bMarkaryan1}
J.Koiller, R.Markarian, S.Q.Kamphorst, and S.P.de Carvalho.
Time-dependent bil\-li\-ards.--- {\it Nonlinearity} {\bf 8}, 983 (1995)

\bibitem{bMarkaryan2}
J.Koiller, R.Markarian, S.Q.Kamphorst, and S.P.de Carvalho.
Static and time-dependent perturbations of the classical elliptical
billiard.--- {\it J. Stat. Phys.} {\bf 83}, 127 (1996).

\bibitem{bMarkaryan3}
R.Markarian. New ergodic billiards: Exact results.---
{\it Nonlinearity} {\bf 6}, 819-841 (1993);

\bibitem{Sinai3}
Ya.G.Sinai. Dynamical system with elastic reflection:
Ergodic properties of dispersing billiards.--- {\it Russ. Math. Surv.}
{\bf 25}, 137-188 (1970).

\bibitem{BnchSinLorenzGas}
L.A.Bunimovich and Ya.G.Sinai. Statistical properties of Lorentz gas
with periodic configuration of scatterers.--- {\it Commun. Math.
Phys.} {\bf 78}, 479-497 (1981).

\bibitem{BnchEgodicProp1}
L.A.Bunimovich. On ergodic properties of some billiards.---
{\it Func. Anal. Appl.} {\bf 8}, 73 (1974)

\bibitem{BnchEgodicProp2}
L.A.Bunimovich. On the ergodic properties of nowhere dispersing
billiards.--- {\it Commun. Math. Phys.} {\bf 65}, 295-312 (1979)

\bibitem{BnchCondStoch}
L.A.Bunimovich. Conditions of stochasticity of two-demensional
billiards.--- {\it Chaos} {\bf 1}, 187-93 (1991).

\bibitem{ZaslKriter}
G.M.Zaslavsiky. Stochasticity in quantum mechanics.---
{\it Phys.Rep.} {\bf 80}, 147-250 (1981)

\bibitem{Fermi}
E.Fermi. On the origin of the cosmic radiation.--- {\it Phys. Rev.}
{\bf 75}, 1169 (1949).

\bibitem{Ulam}
S.M.Ulam. In: {\it Proceedings of the 4th Berkeley
Symp. on Math. Stat. and Pro\-ba\-bi\-lity}.--- University of California
Press {\bf 3}, 315 (1961).

\bibitem{LLBookEn}
A.J.Lichtenberg, M.A.Lieberman. {\it Regular and
Stochastic Motion}.--- (Springer-Ver\-lag, Berlin, 1983).

\bibitem{Levi}
M.Levi. Quasiperiodic motions in superquadratic time-periodic
potentials.--- {\it Com\-mun. Math. Phys.} {\bf 143}, 43-83 (1991).

\bibitem{MyZHTPH}
A.Loskutov, A.B.Ryabov, L.G.Akinshin. Mechanism
of Fermi acceleration in dis\-per\-sing billiards with time-dependent
boundaries.--- {\it J. Exp. Theor. Phys.} {\bf 86}, 966-973 (1999).

\bibitem{MyPHA}
A.Loskutov, A.B.Ryabov and L.G.Akinshin. Properties of some
chaotic billiards with time-dependent boundaries.--- {\it J. Phys. A},
{\bf 44}, 7973--7986 (2000).

\bibitem{bTBPubl}
A.Loskutov, A.B.Ryabov.--- To be published.

\end{thebibliography}
\end{document}